\documentclass[aps,pre,twocolumn,superscriptaddress,showpacs]{revtex4}
\usepackage{graphicx}
\usepackage{color}
\usepackage{amsmath,amssymb}

\begin{document}

\title{Experimental measurement of efficiency and transport
  coherence of a cold atom Brownian motor in optical lattices }

\author{M. Zelan}
\email[]{martin.zelan@physics.umu.se}
\author{H. Hagman}
\author{G. Labaigt}
\affiliation{Department of Physics, Ume{\aa} University, SE-90187 Ume{\aa}, Sweden}
\author{S. Jonsell}
\affiliation{Department of Physics, Stockholm University, SE-10695
  Stockholm, Sweden} 
\author{C. M. Dion}
\affiliation{Department of Physics, Ume{\aa} University, SE-90187 Ume{\aa}, Sweden}

\date{\today}

\begin{abstract}
  The rectification of noise into directed movement or useful energy
  is utilized by many different systems.  The peculiar nature of the
  energy source and conceptual differences between such Brownian motor
  systems makes a characterization of the performance far from
  straightforward. In this work, where the Brownian motor consists of
  atoms interacting with dissipative optical lattices, we adopt
  existing theory and present experimental measurements for both the
  efficiency and the transport coherence. We achieve up to 0.3\% for
  the efficiency and 0.01 for the P\'{e}clet number.
\end{abstract}


\maketitle

Brownian motors (BMs) are devices that can rectify noise into work or
directed motion in the absence of external forces.  They are of
interest for the understanding of fundamental principles in
statistical physics and thermodynamics, and several studies have shown
that they play a crucial part in transport phenomena in nature; see,
for example,
\cite{RonaldDVale04072000,molmotorNature,R.DeanAstumian05091997}.
Since BM's utilize noise, they can work in regions where the inherent
noise is large compared to other interactions. Applications of BM's,
therefore, reach into the nano-scales, where they make ideal tools for
powering up
nano-machines~\cite{MartinG.L.vandenHeuvel07202007,PoweringNanorobots,Man-MadeNanomachines}.
Recent reviews of the subject can be found
in~\cite{RevModPhys.81.387,BM1,BM2,BM3}.

Of particular interest for any motor is the quantification of its
efficiency, usually defined as the ratio of produced work to input
energy.  Due to the peculiar nature of the energy source of BMs,
determination of efficiency is not straightforward.  There have been
several theoretical discussions on the efficiency of
BM's~\cite{Derenyi:1999p1622,PhysRevE.68.021906,PhysRevE.70.061105,r4,RevModPhys.69.1269},
and different performance characteristics have been discussed
in~\cite{linke05}.  We present here experimental measurements of two
performance characteristics of a BM realized with ultracold atoms in
double optical lattices~\cite{OurBM1}: the efficiency, that is, the
fraction of input power driving the directed motion, and the transport
coherence, or the P\'{e}clet number, that is, the comparison between the
drift and the diffusion.  Usually, the efficiency is defined in terms
of the amount of work obtained from the motor against a load.  As no
load is present in our case, we instead follow the
convention~\cite{Derenyi:1999p1622,PhysRevE.68.021906} of defining
``useful energy'' as the energy needed to drive the directed motion of
the atoms against friction.  It has also been argued that including
the dissipation due to friction against the directed motion provides a
better definition of efficiency even when a load is
present~\cite{Derenyi:1999p1622}.

For a BM to be able to function, it has to (i) present an
asymmetry~\cite{Curie} and (ii) be out of thermal
equilibrium~\cite{rat_praw}.  In most cases, the symmetry breaking
arises either from a time-asymmetric periodic driving force with zero
average (rocked ratchet), or by flashing an asymmetric potential
(flashed ratchet).  However, as shown in our system~\cite{OurBM1},
rectification can be achieved by switching between two symmetric
potentials.

The model for the BM used in our experiment was introduced
in~\cite{Fokker1}.  Briefly, particles with mass $m$ move in two
symmetric potentials, $U_1=A_1 \cos(k x)$ and $U_2=A_2 \cos ( k
x+\phi)$, phase shifted by $\phi$, and are randomly transferred
between the two with unequal transfer rates
$\Gamma_{1\rightarrow2}\neq \Gamma_{2\rightarrow1}$.  In addition, the
particles experience a friction force $-\alpha_i \dot{x}$ and a
diffusive force $\xi_i(t)$ in either lattice $i=1,2$. This gives the
equations of motion
\begin{equation}
m\ddot{x} = - \nabla_x U_i(x)-\alpha_i \dot{x} +\xi_i(t) . 
\label{eq:pdot}
\end{equation}
Here, $\xi_i(t)$ satisfies the relations $\langle \xi_i(t)\rangle=0$
and $\langle \xi_i(t)\xi_i(t')\rangle=2 D_i\delta(t-t')$. Thus the
atom is both subject to work from the potential $U_i$ and to
fluctuations and dissipation given by the diffusion coefficient $D_i$
and the friction coefficient $\alpha_i$.

\begin{figure}
\centering
\includegraphics[width=80mm]{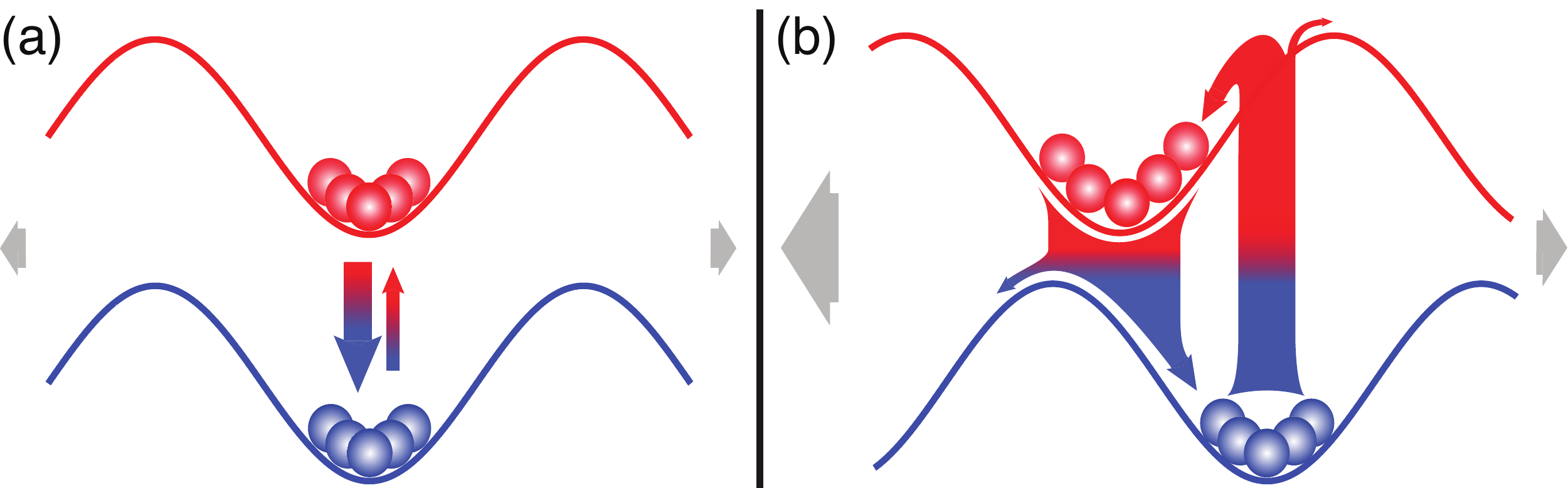}
\caption{(Color online) Schematic drawing of the process driving our
  BM. In each potential, an inherent friction and diffusion is
  present. Vertical arrows indicate the transfer between the
  potentials, horizontal ones the total diffusion. (a) $\phi=0$: atoms
  are transferred between the lattices without any gain in potential
  energy. No symmetry is broken and no drift is induced. (b)
  $\phi=3/2\pi$: the symmetry is broken and the transfer adds energy
  to the system, as a leftward drift.}
\label{Lattice}
\end{figure}

For an atom moving in a single periodic potential, the long-time
average of the work goes to zero, and the atoms reach a steady state
with kinetic temperature $D_i/\alpha_i$.  This changes when it is
transferred between the potentials, changing instantaneously its
potential energy.  The total work on an atom is therefore equal to the
changes in potential energy summed over all jumps between the
potentials.  For identical potentials ($A_1=A_2$, $\phi=0$), no energy
is gained by an atom transferred between the potentials; see
Fig.~\ref{Lattice}(a).
%
%
In this situation, both potentials satisfy the symmetry condition
$U_i(-x)=U_i(x)$, which entails that $\langle \dot{x} \rangle=-\langle
\dot{x} \rangle$, and hence no BM effect is possible~\cite{reimann01}.
Introducing a nonzero $\phi$ between the lattices, the system still
possesses glide reflection~\cite{brown:kanada99}, but because of the
unequal transfer rates between the potentials, there is no symmetry
condition requiring $\langle \dot{x}\rangle=0$, and therefore there
will in general be a rectification~\cite{Curie}.  An exception is the
point $\phi=\pi$, where again $U_i(-x)=U_i(x)$, leading to zero
current, with the input energy gained from the transfer between
potentials only appearing as a heating of the atoms.

Following the discussion in \cite{PhysRevE.68.021906}, we can derive a
detailed balance relation for the input power $P_{\rm in}$ acting on a
particle governed by the equation of motion (\ref{eq:pdot}),
\begin{equation}
  \label{eq:Pin}
  P_{\rm in} =  \frac{{\overline\alpha}\,\overline{p}^2}{m^2} + \frac{\overline{\alpha \delta
      p^2}}{m^2} -
  \frac{N \overline{D}}{m}. 
\end{equation}
Here $\overline{\alpha}$, etc., signifies the time average (over the
two lattice states), and $\delta p(t)$ is the variation of the
momentum $p(t)$ around its long-time average $\overline{p}$. The first
and second terms on the right-hand side of Eq.~(\ref{eq:Pin}) are the
energy loss due to dissipation associated with the directed and the
random motion, respectively. The third term represents the energy
gained through diffusive processes internal to either of the two
lattices, as represented by the constants of momentum diffusion
$D_i$. The factor $N$ is the dimensionality of the system, in our case
$N=3$, and $\delta p^2=\delta p_x^2+\delta p_y^2 +\delta p_z^2\simeq
3\delta p_z^2$. In our system, the temperature might be slightly
different in the different directions~\cite{Jersblad:2003}. However,
this approximation will only have a minor effect on the results.

In~\cite{Derenyi:1999p1622,PhysRevE.68.021906}, the efficiency of a BM
without a load is defined as the energy dissipated by friction acting
against the directed motion over the total energy input, that is,
\begin{equation}
  \label{eq:eta}
  \eta=\frac{\overline{\alpha}\,\overline{p}^2/m}{P_{\rm in}}
=\frac{\overline{\alpha}\,\overline{p}^2/m}{{\overline\alpha}\,\overline{p}^2/m + \overline{\alpha \delta
      p^2}/m - N\overline{D}}.
\end{equation}
Assuming that $\overline{\alpha}$
is uncorrelated to the lattice state, so that
$\overline{\alpha \delta p^2}=\overline{\alpha}\overline{\delta p^2}$,
this expression is simplified to 
\begin{equation}
  \label{eq:eta2}
  \eta
=\frac{\overline{p}^2/m}{\overline{p}^2/m + \overline{\delta
    p^2}/m - N\overline{D}/\overline\alpha}. 
\end{equation}
Simulations indicate that, even for a relatively large difference in
the friction coefficients between the two lattices
($\alpha_2=\alpha_1/2$), this assumption introduces an error in
$P_{\rm in}$ of only about 2\%.
 
The experiment has been described in detail
in~\cite{DOL,OurBM1,setups}.  In short, we use laser cooling to trap
and cool cesium atoms and transfer them into a double optical
lattice~\cite{DOL,setups}.  These are potentials realized from the
interference pattern of laser beams due to a second-order interaction
between the induced atomic dipole moment and the periodic light
fields~\cite{lin-p-lin1}.  The two potentials correspond to two
different hyperfine levels, $F=3$ and $F=4$, of the electronic ground
state of cesium.  Each atom will be transferred between the two
potentials at random times through optical pumping, with rates for
transfer, scattering, and cooling set by the parameters of the laser
fields (intensity and detuning). To collect data, we use absorption
imaging to measure the mean momentum $\overline{p}$ as well as the
size of the atomic cloud. The imaging is done in the horizontal plane
to avoid any effects of gravity~\cite{grav}. To access the mean
momentum spread (the kinetic temperature) $\overline{\delta p^2}$ in
Eq.~(\ref{eq:eta2}), we use a time-of-flight technique that enables
fast and accurate measurements of the distribution of the momentum,
$\delta p_z^2$~\cite{hagman:083109}.

In the experiment we, adjust the potential depths by controlling the
intensities in the lattice beams such that $A_1=A_2$.  To assess the
quantity $\overline{D}/\overline{\alpha}$, we study the system at
$\phi=0$, where the potentials are identical and there is no BM effect
since the transfer between the potentials does not change the energy
of the system ($P_{\rm in}=0$).  From the energy balance
(\ref{eq:Pin}), we then obtain, in agreement with the equipartition
theorem,
\begin{equation}
  \label{eq:D0}
 \left.\overline{E_{\rm kin}}\right|_{\phi=0}=\left.\frac{\overline{\delta p^2}}{2m}\right|_{\phi=0}=\frac{N}{2}\frac{{\overline  D}}{{\overline
      \alpha}}.
\end{equation}
The association of $\overline{D}/\overline{\alpha}$ with the kinetic
temperature at $\phi=0$ assumes that the diffusion and friction
constants are independent of the relative phase of the lattices.
Following the standard model of Sisyphus cooling~\cite{cool:nobel98},
our model (\ref{eq:pdot}) assumes that diffusion and friction are
spatially homogeneous. It should be noted, though, that in a more
accurate model these coefficients are dependent on the position $x$ of
the atom in the lattice. The spatial distribution of atoms in either
lattice will have some dependence on $\phi$, which will translate into
a dependence on the spatially averaged friction and diffusion
coefficients.  This is ignored in our model, introducing a degree of
approximation in Eq.~(\ref{eq:eta2}) for efficiency.

\begin{figure}
\centering
\includegraphics[width=80mm]{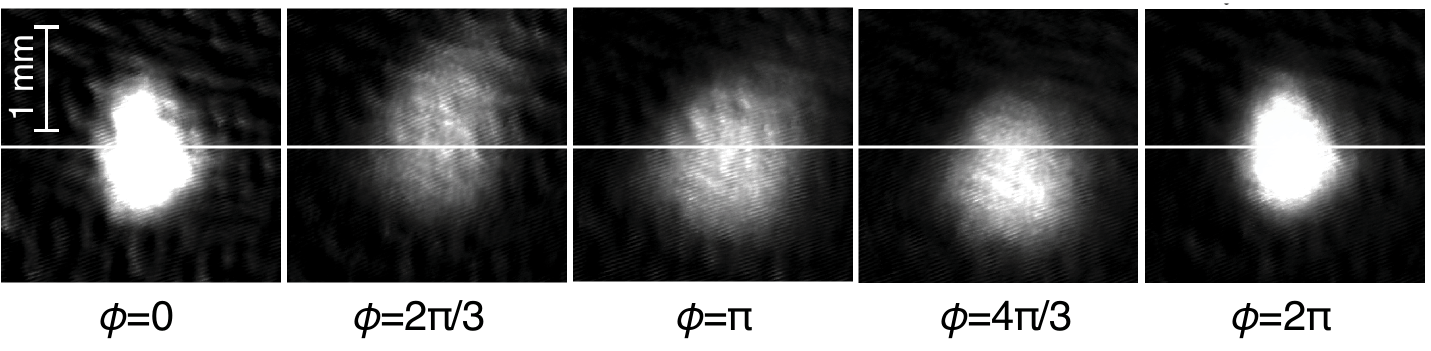}
\caption{Absorption images of atoms in the double optical lattice for
  five different relative spatial phases where the potential depth is
  200 $\mu \rm{K}$. The atoms are kept in the lattices for 150 ms. For
  $\phi=0$ and $2\pi$, the drift is zero and the diffusion is
  small. For $\phi=\pi$, the drift is also zero, but the diffusion is
  much larger. Images 2 and 4 show maximum drifts in the two different
  directions.}
\label{Fig_absimg}
\end{figure}

\begin{figure}
\centering
\includegraphics[width=80mm]{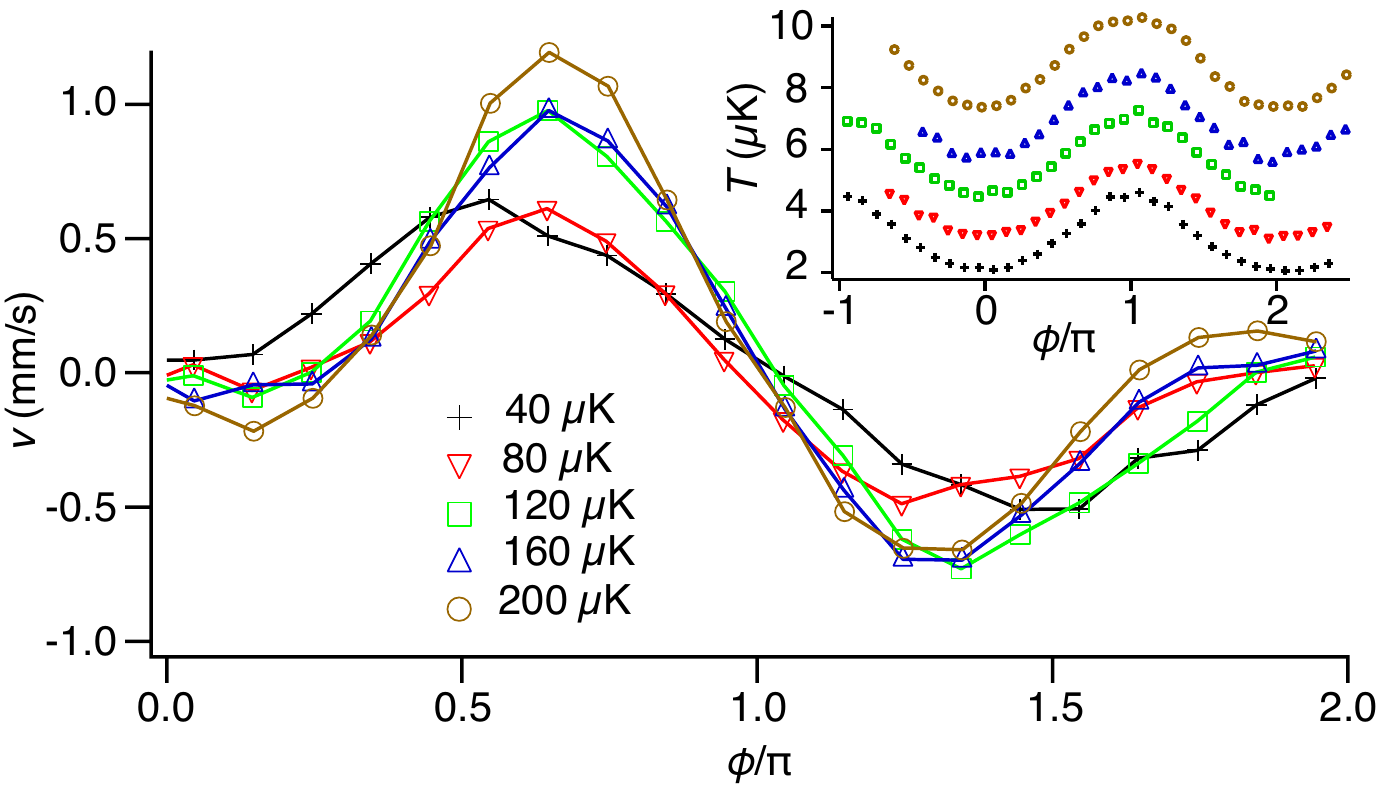}
\caption{(Color online) The drift velocity versus the phase shift for
  a lattice holding time of 150~ms. Inset: the kinetic temperature for
  the same data.}
\label{drift}
\end{figure}

\begin{figure}
\centering
\includegraphics[width=80mm]{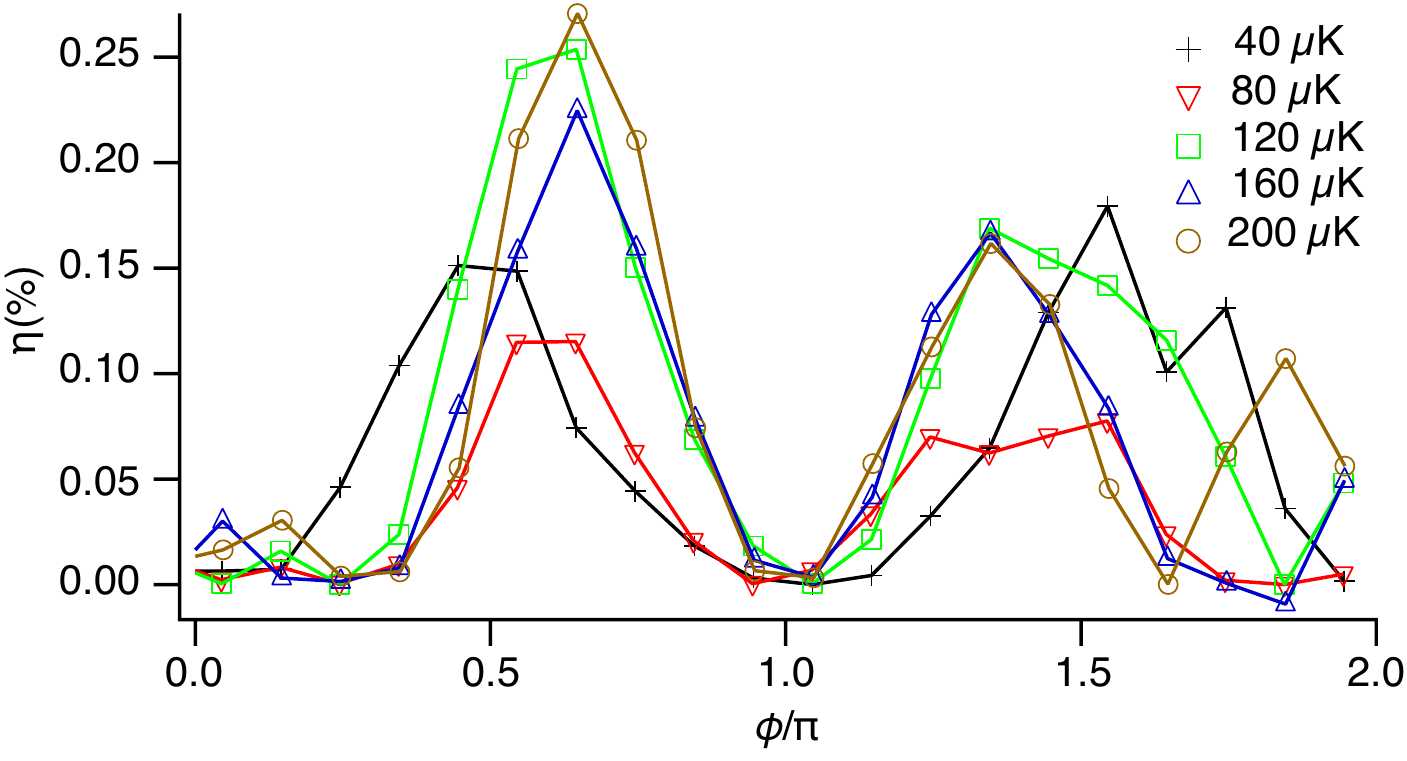}
\caption{ (Color online) The efficiency, according to
  Eq.~(\ref{eq:eta2}), of the Brownian motor as a function of the
  relative spatial phase for five different potential depths. The
  greatest efficiency is achieved for $\phi$ equal to $2\pi/3$ and
  $4\pi/3$ and drops to zero for $\phi=\pi$.}
\label{efficiency}
\end{figure}

Absorption images were taken for five different potential depths.  In
Fig.~\ref{Fig_absimg}, we show typical raw data for atoms kept 150~ms
in the lattices.  A clear drift is seen in images 2 and 4, while
images 1, 3, and 5 show no drift, as expected.
Images such as Fig.~\ref{Fig_absimg} have been taken for $0 \leq \phi
\leq 2\pi$ for potential depths between $40$ and $200\ \mu \rm{K}$,
and the analyzed data can be seen in Fig.~\ref{drift} in terms of the
drift velocity.
The induced drifts are expected to be symmetric around
$\phi=\pi$. However, slightly larger drifts are observed for
$\phi=2\pi/3$ than for $\phi=4\pi/3$, most likely due to experimental
limitations in the alignment and the intensity balance of the lattice
beams. Also shown in Fig.~\ref{drift} is the kinetic temperature for
the same parameters. We find that the baseline of the kinetic
temperature increases with the potential depth, while the amplitude of
its variation with $\phi$ is roughly unchanged. As discussed earlier,
the $\phi$-dependent kinetic temperature is represented by the second
term on the right-hand side of Eq.~(\ref{eq:Pin}), while the baseline
(or kinetic temperature at $\phi$=0) is represented by the third term.
Hence their difference is the variation of the kinetic temperature
with $\phi$.  Our data show that this variation is approximately the
same for different potential depths.  Hence, the greater efficiency
for larger potential depths is mainly due to the increase in the drift
momentum.  Using the data from Fig.~\ref{drift} in
Eq.~(\ref{eq:eta2}), we obtain the efficiency as a function of $\phi$,
where the maximum efficiency is close to 0.3\%; see
Fig.~\ref{efficiency}.

An alternative way to characterize the rectified motion is by the
coherence of the transport, where the linear transport is compared to
the diffusion.  This can be quantified using the P\'{e}clet
number~\cite{linke05},
\begin{equation}
\text{Pe} \equiv \frac{|\langle v \rangle | l }{\tilde{D}_{\rm eff}},
\label{eff_pec}
\end{equation}
where $l$ is a characteristic length of the system, in our case the
lattice constant, and $\tilde{D}_{\rm eff}$ is the effective spatial
diffusion given by
\begin{equation}
  \tilde{D}_{\rm eff} \equiv \lim_{t \to +\infty} \frac{\langle x^2(t)\rangle-\langle x(t)\rangle ^2}{2t}.
\label{effnoload}
\end{equation}
For atoms in dissipative optical lattices, where thermal fluctuations
play an important role, $\tilde{D}_{\rm eff}$ becomes the spatial
diffusion constant $\tilde{D}=\langle [\delta x(t)-\delta x(0)]^2
\rangle/(2t)$, where $\delta x(t)=x(t)-\langle x(t)\rangle$
\cite{Machura:2005p5573}. This quantity can be calculated from the
expansion of the atomic cloud in the optical lattices, where the size
of the cloud is given by
\begin{equation}
\sigma_t=\sqrt{\sigma_{0}{^2}+2\tilde{D}t},
\label{size}
\end{equation}
%
with $\sigma_t$ the root-mean-square radius at time $t$.

\begin{figure}
\centering
\includegraphics[width=80mm]{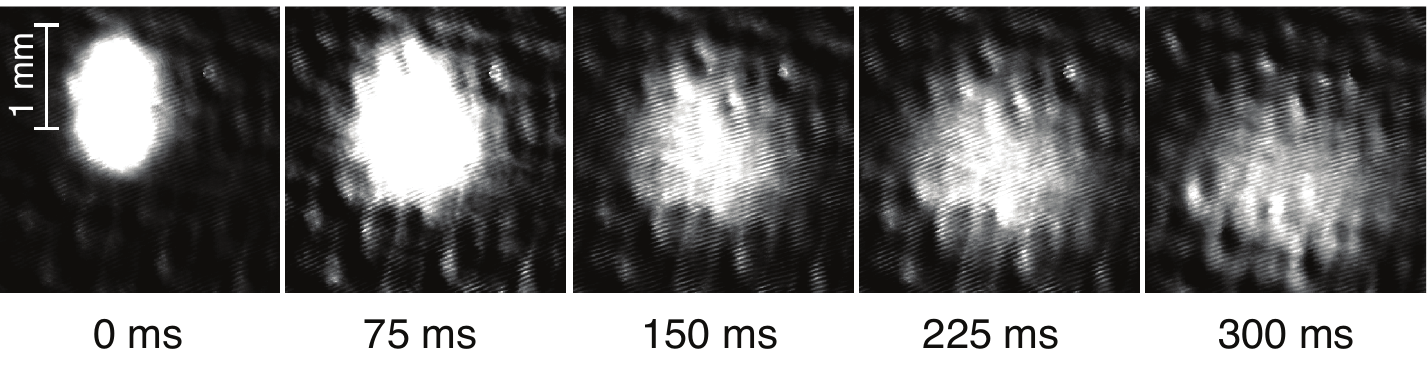}
\caption{Time evolution of the atomic sample for a potential depth of
  $200\ \mu\mathrm{K}$.  Both the drift and the diffusion are seen.}
\label{Fig_5}
\end{figure}

In order to quantify the performance in terms of the P\'{e}clet number,
series of absorption images of the time evolution of the atomic cloud,
such as shown in Fig.~\ref{Fig_5}, have been taken. The phase is set
to achieve maximum drift ($\phi=2\pi/3$).

\begin{figure}
\centering
\includegraphics[width=80mm]{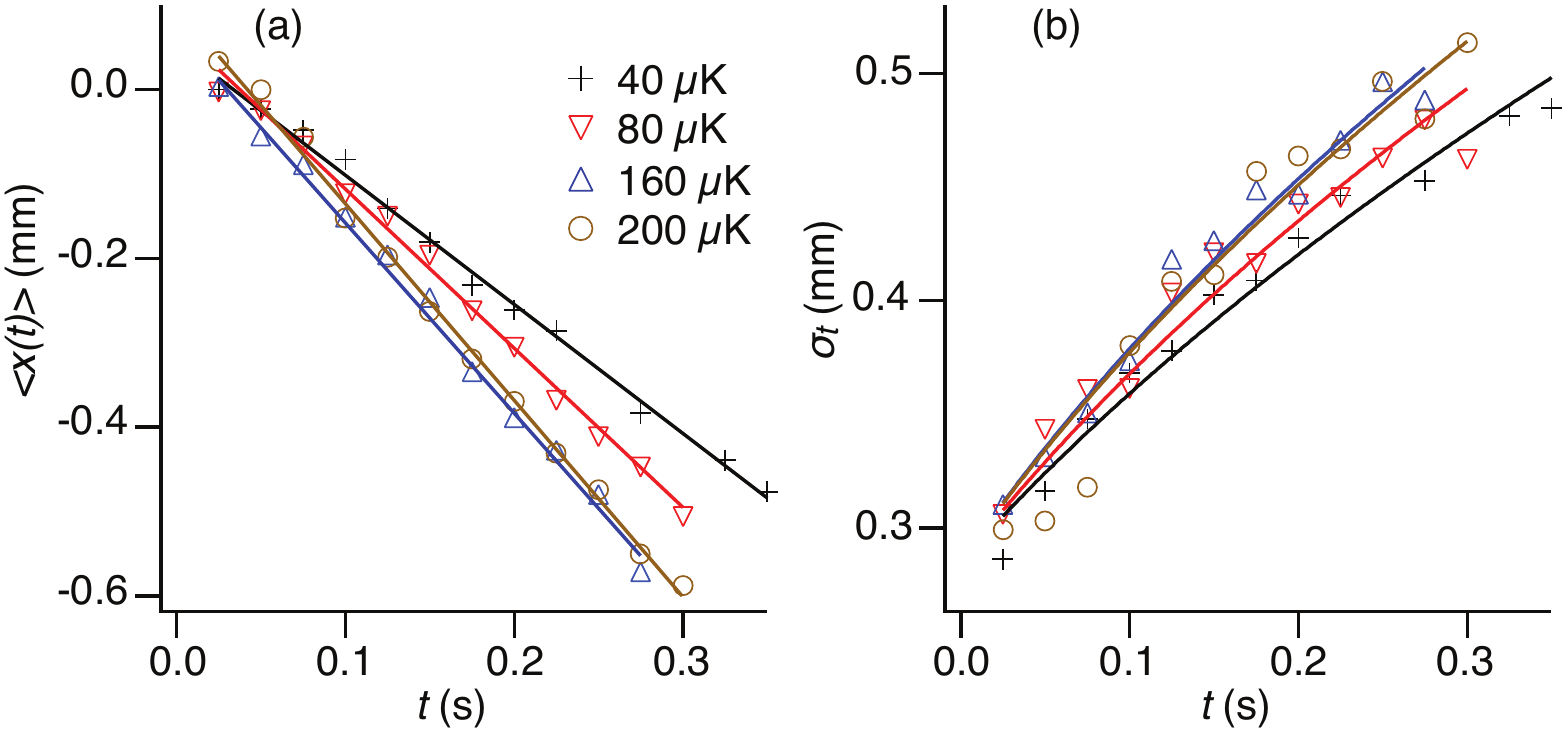}
\caption{(Color online) (a) Position of the center of mass of the
  atomic sample. (b) Root-mean-square radius of the atomic sample as a
  function of holding time for four different potential depths.  The
  center of mass moves linearly as expected~\cite{OurBM1}, and
  indicates faster drifts for higher potential depths. The size of the
  cloud grows with time due to diffusion according to
  Eq.~(\ref{size}).}
\label{Fig_6}
\end{figure}

\begin{figure}
\centering
\includegraphics[width=80mm]{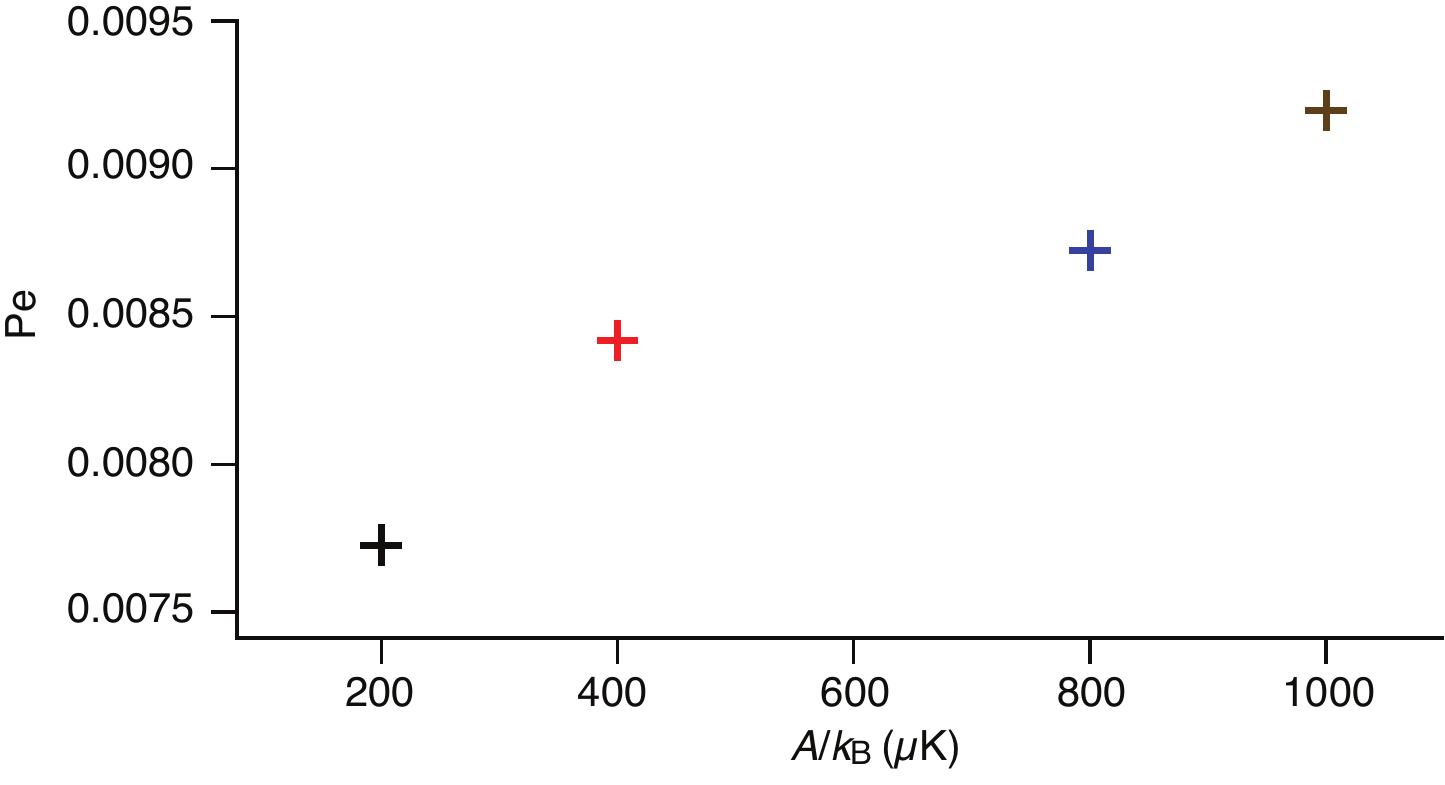}
\caption{(Color online) The measured P\'{e}clet number,
  Eq.~(\ref{eff_pec}), as a function of potential depth. The data
  indicates a greater coherence in the transport for higher potential
  depths.}
\label{peclet}
\end{figure}

In Fig.~\ref{Fig_6}(a), a series of such images have been analyzed and
the drift is plotted against the holding time in the lattice.
In Fig.~\ref{Fig_6}(b), the width of the sample is shown against the
holding time, from which the diffusion constant $\tilde{D}_{\rm eff}$
can be extracted by fitting to Eq.~(\ref{size}). Combining the result
with the measured average velocity of the sample, according to
Eq.~(\ref{eff_pec}), gives the P\'{e}clet number for different potential
depths; see Fig.~\ref{peclet}.

In conclusion, we have adopted existing theory and presented
experimental results for two measures of the performance of a Brownian
motor, namely the efficiency and the P\'{e}clet number, in a system of
ultracold atoms in a double optical lattice.  The results indicate
trends that give higher efficiency and transport coherence for deeper
potentials, and are in agreement with the values of the P\'{e}clet number
that were predicted for similar systems~\cite{PhysRevA.81.013416}.
Although our BM prototype differs from other BM's, the fundamental
principles are the same, and hence these kinds of characteristic
measurements allow for interesting comparisons between BM systems from
different fields.

\begin{acknowledgments}
This project was supported by the Swedish Research Council, Knut \&
Alice Wallenbergs Stiftelse, Carl Trygger Stiftelse, and Ume{\aa} University.
\end{acknowledgments}


\begin{thebibliography}{31}
\expandafter\ifx\csname natexlab\endcsname\relax\def\natexlab#1{#1}\fi
\expandafter\ifx\csname bibnamefont\endcsname\relax
  \def\bibnamefont#1{#1}\fi
\expandafter\ifx\csname bibfnamefont\endcsname\relax
  \def\bibfnamefont#1{#1}\fi
\expandafter\ifx\csname citenamefont\endcsname\relax
  \def\citenamefont#1{#1}\fi
\expandafter\ifx\csname url\endcsname\relax
  \def\url#1{\texttt{#1}}\fi
\expandafter\ifx\csname urlprefix\endcsname\relax\def\urlprefix{URL }\fi
\providecommand{\bibinfo}[2]{#2}
\providecommand{\eprint}[2][]{\url{#2}}

\bibitem[{\citenamefont{Vale and Milligan}(2000)}]{RonaldDVale04072000}
\bibinfo{author}{\bibfnamefont{R.~D.} \bibnamefont{Vale}} \bibnamefont{and}
  \bibinfo{author}{\bibfnamefont{R.~A.} \bibnamefont{Milligan}},
  \bibinfo{journal}{Science} \textbf{\bibinfo{volume}{288}},
  \bibinfo{pages}{88} (\bibinfo{year}{2000}).

\bibitem[{\citenamefont{Schliwa and Woehlke}(2003)}]{molmotorNature}
\bibinfo{author}{\bibfnamefont{M.}~\bibnamefont{Schliwa}} \bibnamefont{and}
  \bibinfo{author}{\bibfnamefont{G.}~\bibnamefont{Woehlke}},
  \bibinfo{journal}{Nature} \textbf{\bibinfo{volume}{422}},
  \bibinfo{pages}{759} (\bibinfo{year}{2003}).

\bibitem[{\citenamefont{Astumian}(1997)}]{R.DeanAstumian05091997}
\bibinfo{author}{\bibfnamefont{R.~D.} \bibnamefont{Astumian}},
  \bibinfo{journal}{Science} \textbf{\bibinfo{volume}{276}},
  \bibinfo{pages}{917} (\bibinfo{year}{1997}).

\bibitem[{\citenamefont{van~den Heuvel and
  Dekker}(2007)}]{MartinG.L.vandenHeuvel07202007}
\bibinfo{author}{\bibfnamefont{M.~G.~L.} \bibnamefont{van~den Heuvel}}
  \bibnamefont{and} \bibinfo{author}{\bibfnamefont{C.}~\bibnamefont{Dekker}},
  \bibinfo{journal}{Science} \textbf{\bibinfo{volume}{317}},
  \bibinfo{pages}{333} (\bibinfo{year}{2007}).

\bibitem[{\citenamefont{Mallouk and Sen}(2009)}]{PoweringNanorobots}
\bibinfo{author}{\bibfnamefont{T.~E.} \bibnamefont{Mallouk}} \bibnamefont{and}
  \bibinfo{author}{\bibfnamefont{A.}~\bibnamefont{Sen}}, \bibinfo{journal}{Sci.
  Am.} \textbf{\bibinfo{volume}{300}}, \bibinfo{pages}{72}
  (\bibinfo{year}{2009}).

\bibitem[{\citenamefont{Wang}(2009)}]{Man-MadeNanomachines}
\bibinfo{author}{\bibfnamefont{J.}~\bibnamefont{Wang}}, \bibinfo{journal}{ACS
  Nano} \textbf{\bibinfo{volume}{3}}, \bibinfo{pages}{4}
  (\bibinfo{year}{2009}).

\bibitem[{\citenamefont{H\"{a}nggi and Marchesoni}(2009)}]{RevModPhys.81.387}
\bibinfo{author}{\bibfnamefont{P.}~\bibnamefont{H\"{a}nggi}} \bibnamefont{and}
  \bibinfo{author}{\bibfnamefont{F.}~\bibnamefont{Marchesoni}},
  \bibinfo{journal}{Rev. Mod. Phys.} \textbf{\bibinfo{volume}{81}},
  \bibinfo{pages}{387} (\bibinfo{year}{2009}).

\bibitem[{\citenamefont{H\"{a}nggi et~al.}(2005)\citenamefont{H\"{a}nggi,
  Marchesoni, and Nori}}]{BM1}
\bibinfo{author}{\bibfnamefont{P.}~\bibnamefont{H\"{a}nggi}},
  \bibinfo{author}{\bibfnamefont{F.}~\bibnamefont{Marchesoni}},
  \bibnamefont{and} \bibinfo{author}{\bibfnamefont{F.}~\bibnamefont{Nori}},
  \bibinfo{journal}{Ann. Phys. (Leipzig)} \textbf{\bibinfo{volume}{14}},
  \bibinfo{pages}{51} (\bibinfo{year}{2005}).

\bibitem[{\citenamefont{Reimann}(2002)}]{BM2}
\bibinfo{author}{\bibfnamefont{P.}~\bibnamefont{Reimann}},
  \bibinfo{journal}{Phys. Rep.} \textbf{\bibinfo{volume}{361}},
  \bibinfo{pages}{57} (\bibinfo{year}{2002}).

\bibitem[{\citenamefont{Astumian and H\"{a}nggi}(2002)}]{BM3}
  \bibinfo{author}{\bibfnamefont{R.~D.} \bibnamefont{Astumian}}
  \bibnamefont{and}
  \bibinfo{author}{\bibfnamefont{P.}~\bibnamefont{H\"{a}nggi}},
  \bibinfo{journal}{Physics Today} \textbf{\bibinfo{volume}{55}} (11),
  \bibinfo{pages}{33} (\bibinfo{year}{2002}).

\bibitem[{\citenamefont{Der\'{e}nyi et~al.}(1999)\citenamefont{Der\'{e}nyi,
  Bier, and Astumian}}]{Derenyi:1999p1622}
\bibinfo{author}{\bibfnamefont{I.}~\bibnamefont{Der\'{e}nyi}},
  \bibinfo{author}{\bibfnamefont{M.}~\bibnamefont{Bier}}, \bibnamefont{and}
  \bibinfo{author}{\bibfnamefont{R.~D.} \bibnamefont{Astumian}},
  \bibinfo{journal}{Phys. Rev. Lett.} \textbf{\bibinfo{volume}{83}},
  \bibinfo{pages}{903} (\bibinfo{year}{1999}).

\bibitem[{\citenamefont{Suzuki and Munakata}(2003)}]{PhysRevE.68.021906}
\bibinfo{author}{\bibfnamefont{D.}~\bibnamefont{Suzuki}} \bibnamefont{and}
  \bibinfo{author}{\bibfnamefont{T.}~\bibnamefont{Munakata}},
  \bibinfo{journal}{Phys. Rev. E} \textbf{\bibinfo{volume}{68}},
  \bibinfo{pages}{021906} (\bibinfo{year}{2003}).

\bibitem[{\citenamefont{Machura et~al.}(2004)\citenamefont{Machura, Kostur,
  Talkner, \L{}uczka, Marchesoni, and H\"{a}nggi}}]{PhysRevE.70.061105}
\bibinfo{author}{\bibfnamefont{L.}~\bibnamefont{Machura}},
  \bibinfo{author}{\bibfnamefont{M.}~\bibnamefont{Kostur}},
  \bibinfo{author}{\bibfnamefont{P.}~\bibnamefont{Talkner}},
  \bibinfo{author}{\bibfnamefont{J.}~\bibnamefont{\L{}uczka}},
  \bibinfo{author}{\bibfnamefont{F.}~\bibnamefont{Marchesoni}},
  \bibnamefont{and} \bibinfo{author}{\bibfnamefont{P.}~\bibnamefont{H\"{a}nggi}},
  \bibinfo{journal}{Phys. Rev. E} \textbf{\bibinfo{volume}{70}},
  \bibinfo{pages}{061105} (\bibinfo{year}{2004}).

\bibitem[{\citenamefont{Reimann and H\"{a}nggi}(2002)}]{r4}
\bibinfo{author}{\bibfnamefont{P.}~\bibnamefont{Reimann}} \bibnamefont{and}
  \bibinfo{author}{\bibfnamefont{P.}~\bibnamefont{H\"{a}nggi}},
  \bibinfo{journal}{Appl. Phys. A} \textbf{\bibinfo{volume}{75}},
  \bibinfo{pages}{169} (\bibinfo{year}{2002}).

\bibitem[{\citenamefont{J\"{u}licher et~al.}(1997)\citenamefont{J\"{u}licher,
  Ajdari, and Prost}}]{RevModPhys.69.1269}
\bibinfo{author}{\bibfnamefont{F.}~\bibnamefont{J\"{u}licher}},
  \bibinfo{author}{\bibfnamefont{A.}~\bibnamefont{Ajdari}}, \bibnamefont{and}
  \bibinfo{author}{\bibfnamefont{J.}~\bibnamefont{Prost}},
  \bibinfo{journal}{Rev. Mod. Phys.} \textbf{\bibinfo{volume}{69}},
  \bibinfo{pages}{1269} (\bibinfo{year}{1997}).

\bibitem[{\citenamefont{Linke et~al.}(2005)\citenamefont{Linke, Downton, and
  Zuckermann}}]{linke05}
\bibinfo{author}{\bibfnamefont{H.}~\bibnamefont{Linke}},
  \bibinfo{author}{\bibfnamefont{M.~T.} \bibnamefont{Downton}},
  \bibnamefont{and} \bibinfo{author}{\bibfnamefont{M.~J.}
  \bibnamefont{Zuckermann}}, \bibinfo{journal}{Chaos}
  \textbf{\bibinfo{volume}{15}}, \bibinfo{pages}{026111}
  (\bibinfo{year}{2005}).

\bibitem[{\citenamefont{Sj\"{o}lund et~al.}(2006)\citenamefont{Sj\"{o}lund, Petra,
  Dion, Jonsell, Nyl\'{e}n, Sanchez-Palencia, and Kastberg}}]{OurBM1}
\bibinfo{author}{\bibfnamefont{P.}~\bibnamefont{Sj\"{o}lund}},
  \bibinfo{author}{\bibfnamefont{S.~J.~H.} \bibnamefont{Petra}},
  \bibinfo{author}{\bibfnamefont{C.~M.} \bibnamefont{Dion}},
  \bibinfo{author}{\bibfnamefont{S.}~\bibnamefont{Jonsell}},
  \bibinfo{author}{\bibfnamefont{M.}~\bibnamefont{Nyl\'{e}n}},
  \bibinfo{author}{\bibfnamefont{L.}~\bibnamefont{Sanchez-Palencia}},
  \bibnamefont{and} \bibinfo{author}{\bibfnamefont{A.}~\bibnamefont{Kastberg}},
  \bibinfo{journal}{Phys. Rev. Lett.} \textbf{\bibinfo{volume}{96}},
  \bibinfo{pages}{190602} (\bibinfo{year}{2006}).

\bibitem[{\citenamefont{Curie}(1894)}]{Curie}
\bibinfo{author}{\bibfnamefont{P.}~\bibnamefont{Curie}}, \bibinfo{journal}{J.
  Phys. III (Paris)} \textbf{\bibinfo{volume}{3}}, \bibinfo{pages}{393}
  (\bibinfo{year}{1894}).

\bibitem[{\citenamefont{Feynman et~al.}(1963)\citenamefont{Feynman, Lieghton,
  and Sands}}]{rat_praw}
\bibinfo{author}{\bibfnamefont{R.~P.} \bibnamefont{Feynman}},
  \bibinfo{author}{\bibfnamefont{R.~B.} \bibnamefont{Lieghton}},
  \bibnamefont{and} \bibinfo{author}{\bibfnamefont{M.}~\bibnamefont{Sands}},
  \emph{\bibinfo{title}{The Feynman Lectures on Physics}},
  vol.~\bibinfo{volume}{1} (\bibinfo{publisher}{Addison-Wesley},
  \bibinfo{address}{Reading, MA}, \bibinfo{year}{1963}).

\bibitem[{\citenamefont{Sanchez-Palencia}(2004)}]{Fokker1}
\bibinfo{author}{\bibfnamefont{L.}~\bibnamefont{Sanchez-Palencia}},
  \bibinfo{journal}{Phys. Rev. E} \textbf{\bibinfo{volume}{70}},
  \bibinfo{pages}{011102} (\bibinfo{year}{2004}).

\bibitem[{\citenamefont{Reimann}(2001)}]{reimann01}
\bibinfo{author}{\bibfnamefont{P.}~\bibnamefont{Reimann}},
  \bibinfo{journal}{Phys. Rev. Lett.} \textbf{\bibinfo{volume}{86}},
  \bibinfo{pages}{4992} (\bibinfo{year}{2001}).

\bibitem[{\citenamefont{Kanada and Sasaki}(1999)}]{brown:kanada99}
\bibinfo{author}{\bibfnamefont{R.}~\bibnamefont{Kanada}} \bibnamefont{and}
  \bibinfo{author}{\bibfnamefont{K.}~\bibnamefont{Sasaki}},
  \bibinfo{journal}{J. Phys. Soc. Japan} \textbf{\bibinfo{volume}{68}},
  \bibinfo{pages}{3759} (\bibinfo{year}{1999}).

\bibitem[{\citenamefont{Jersblad et~al.}(2003)\citenamefont{Jersblad, Ellmann,
  and Kastberg}}]{Jersblad:2003}
\bibinfo{author}{\bibfnamefont{J.}~\bibnamefont{Jersblad}},
  \bibinfo{author}{\bibfnamefont{H.}~\bibnamefont{Ellmann}}, \bibnamefont{and}
  \bibinfo{author}{\bibfnamefont{A.}~\bibnamefont{Kastberg}},
  \bibinfo{journal}{Eur. Phys. J. D} \textbf{\bibinfo{volume}{22}},
  \bibinfo{pages}{333} (\bibinfo{year}{2003}).

\bibitem[{\citenamefont{Ellmann
  et~al.}(2003{\natexlab{a}})\citenamefont{Ellmann, Jersblad, and
  Kastberg}}]{DOL}
\bibinfo{author}{\bibfnamefont{H.}~\bibnamefont{Ellmann}},
  \bibinfo{author}{\bibfnamefont{J.}~\bibnamefont{Jersblad}}, \bibnamefont{and}
  \bibinfo{author}{\bibfnamefont{A.}~\bibnamefont{Kastberg}},
  \bibinfo{journal}{Phys. Rev. Lett.} \textbf{\bibinfo{volume}{90}},
  \bibinfo{pages}{053001} (\bibinfo{year}{2003}{\natexlab{a}}).

\bibitem[{\citenamefont{Ellmann
  et~al.}(2003{\natexlab{b}})\citenamefont{Ellmann, Jersblad, and
  Kastberg}}]{setups}
\bibinfo{author}{\bibfnamefont{H.}~\bibnamefont{Ellmann}},
  \bibinfo{author}{\bibfnamefont{J.}~\bibnamefont{Jersblad}}, \bibnamefont{and}
  \bibinfo{author}{\bibfnamefont{A.}~\bibnamefont{Kastberg}},
  \bibinfo{journal}{Eur. Phys. J.~D} \textbf{\bibinfo{volume}{22}},
  \bibinfo{pages}{355} (\bibinfo{year}{2003}{\natexlab{b}}).

\bibitem[{\citenamefont{Grynberg and Mennerat-Robillard}(2001)}]{lin-p-lin1}
\bibinfo{author}{\bibfnamefont{G.}~\bibnamefont{Grynberg}} \bibnamefont{and}
  \bibinfo{author}{\bibfnamefont{C.}~\bibnamefont{Mennerat-Robillard}},
  \bibinfo{journal}{Phys. Rep.} \textbf{\bibinfo{volume}{355}},
  \bibinfo{pages}{35} (\bibinfo{year}{2001}).

\bibitem[{\citenamefont{Zelan et~al.}(2010)\citenamefont{Zelan, Hagman,
  Karlsson, Dion, and Kastberg}}]{grav}
\bibinfo{author}{\bibfnamefont{M.}~\bibnamefont{Zelan}},
  \bibinfo{author}{\bibfnamefont{H.}~\bibnamefont{Hagman}},
  \bibinfo{author}{\bibfnamefont{K.}~\bibnamefont{Karlsson}},
  \bibinfo{author}{\bibfnamefont{C.~M.} \bibnamefont{Dion}}, \bibnamefont{and}
  \bibinfo{author}{\bibfnamefont{A.}~\bibnamefont{Kastberg}},
  \bibinfo{journal}{Phys. Rev. E} \textbf{\bibinfo{volume}{82}},
  \bibinfo{pages}{031136} (\bibinfo{year}{2010}).

\bibitem[{\citenamefont{Hagman et~al.}(2009)\citenamefont{Hagman, Sj\"{o}lund,
  Petra, Nyl\'{e}n, Kastberg, Ellmann, and Jersblad}}]{hagman:083109}
\bibinfo{author}{\bibfnamefont{H.}~\bibnamefont{Hagman}},
  \bibinfo{author}{\bibfnamefont{P.}~\bibnamefont{Sj\"{o}lund}},
  \bibinfo{author}{\bibfnamefont{S.~J.~H.} \bibnamefont{Petra}},
  \bibinfo{author}{\bibfnamefont{M.}~\bibnamefont{Nyl\'{e}n}},
  \bibinfo{author}{\bibfnamefont{A.}~\bibnamefont{Kastberg}},
  \bibinfo{author}{\bibfnamefont{H.}~\bibnamefont{Ellmann}}, \bibnamefont{and}
  \bibinfo{author}{\bibfnamefont{J.}~\bibnamefont{Jersblad}},
  \bibinfo{journal}{J. Appl. Phys.} \textbf{\bibinfo{volume}{105}},
  \bibinfo{eid}{083109} (\bibinfo{year}{2009}).

\bibitem[{coo()}]{cool:nobel98}
\bibinfo{note}{S. Chu, Rev.\ Mod.\ Phys.\ \textbf{70}, 685 (1998); C.
  Cohen-Tannoudji, \emph{ibid.} \textbf{70}, 707 (1998); W. D. Phillips,
  \emph{ibid.} \textbf{70}, 721 (1998)}.

\bibitem[{\citenamefont{Machura et~al.}(2005)\citenamefont{Machura, Kostur,
  Marchesoni, Talkner, H\"{a}nggi, and {\L}uczka}}]{Machura:2005p5573}
\bibinfo{author}{\bibfnamefont{L.}~\bibnamefont{Machura}},
  \bibinfo{author}{\bibfnamefont{M.}~\bibnamefont{Kostur}},
  \bibinfo{author}{\bibfnamefont{F.}~\bibnamefont{Marchesoni}},
  \bibinfo{author}{\bibfnamefont{P.}~\bibnamefont{Talkner}},
  \bibinfo{author}{\bibfnamefont{P.}~\bibnamefont{H\"{a}nggi}}, \bibnamefont{and}
  \bibinfo{author}{\bibfnamefont{J.}~\bibnamefont{{\L}uczka}},
  \bibinfo{journal}{J. Phys.: Cond. Mat.} \textbf{\bibinfo{volume}{17}},
  \bibinfo{pages}{S3741} (\bibinfo{year}{2005}).

\bibitem[{\citenamefont{Kolton and Renzoni}(2010)}]{PhysRevA.81.013416}
\bibinfo{author}{\bibfnamefont{A.~B.} \bibnamefont{Kolton}} \bibnamefont{and}
  \bibinfo{author}{\bibfnamefont{F.}~\bibnamefont{Renzoni}},
  \bibinfo{journal}{Phys. Rev. A} \textbf{\bibinfo{volume}{81}},
  \bibinfo{pages}{013416} (\bibinfo{year}{2010}).

\end{thebibliography}

\end{document}